\documentclass[12pt,preprint]{aastex}

\usepackage{amsmath}
\usepackage{natbib}
\usepackage{graphicx}
\usepackage{color}
\usepackage[breaklinks,colorlinks,citecolor=blue]{hyperref}
\usepackage{epsfig}
\usepackage{epstopdf}

\usepackage{mathrsfs}

\begin{document}
\title{Thermodynamics and phase transition in rotational Kiselev black hole}

\author{
  Zhaoyi Xu,\altaffilmark{1,2,3}
  Yi Liao\altaffilmark{4,5}
 and Jiancheng Wang,\altaffilmark{1,2,3}
 }

\altaffiltext{1}{Yunnan Observatories, Chinese Academy of Sciences, 396 Yangfangwang, Guandu District, Kunming, 650216, P. R. China; {\tt zyxu88@ynao.ac.cn,jcwang@ynao.ac.cn}}
\altaffiltext{2}{University of Chinese Academy of Sciences, Beijing, 100049, P. R. China}
\altaffiltext{3}{Key Laboratory for the Structure and Evolution of Celestial Objects, Chinese Academy of Sciences, 396 Yangfangwang, Guandu District, Kunming, 650216, P. R. China}
\altaffiltext{4}{Department of Physics, National University of Defense Technology, Changsha, 410073, P. R. China; {\tt liaoyitianyi@gmail.com}}
\altaffiltext{5} {Interdisciplinary Center for Quantum Information, National University of Defense Technology, Changsha, 410073, P. R. China}

\shorttitle{Thermodynamics and Phase Transition in Rotational Kiselev Black Hole}
\shortauthors{Z Y. Xu and J C.  Wang}

\begin{abstract}
In this work, we investigate the thermodynamic properties of rotational Kiselev black holes (KBH). Specifically, we use the first-order approximation of the event horizon (EH) to calculate thermodynamic properties for general equations of state $\omega$. These thermodynamic properties include areas, entropies, horizon radii, surface gravities, surface temperatures, Komar energies and irreducible masses at the Cauchy horizon (CH) and EH. We study the products of these thermodynamic quantities, we find that these products are determined by the equation of state $\omega$ and strength parameter $\alpha$.
In the case of the quintessence matter ($\omega=-2/3$), radiation ($\omega=1/3$) and dust ($\omega=0$), we discuss their properties in detail. We also generalize the Smarr mass formula and Christodoulou-Ruffini mass formula to rotational KBH. Finally we study the phase transition and thermodynamic geometry for rotational KBH with radiation ($\omega=1/3$). Through analysis, we find that this phase transition is a second order phase transition. Furthermore, we also obtain the scalar curvature in the thermodynamic geometry framework, indicating that the radiation matter may change the phase transition condition and properties for Kerr black hole.
\end{abstract}


\section{INTRODUCTION}
Black holes are important objects in physics, especially for quantum gravity. They have the analogy of laws between dynamics and thermodynamics e.g., (\cite{1973CMaPh..31..161B,1972NCimL...4..737B,1974PhRvD...9.3292B,1973PhRvD...7.2333B,1970PhRvL...1596,1971PhRvD...4.3552C,1974Natur.248...30H,1971PhRvL..26.1344H,1971NPhS..229..177P}). Related research has been summarized as black hole thermodynamics. In this framework, the temperature and entropy are analogous to the surface gravity and area of black hole. The laws of thermodynamics in black hole have been discussed by many authers e.g., (\cite{2007PhLB..656..217A,2011GReGr..43.1061J,2011IJTP...50..465J,2011Ap&SS.335..339J}). The ¡®universal property¡¯ of horizon entropy sum of in AdS black hole  spacetime are discussed (\cite{2014JHEP...01..031W}), the Kerr-AdS (K-AdS) black hole and Kerr-Newman-AdS (KN-AdS) black holes situation also have been studied e.g.,(\cite{2010JHEP...04..118S}and its references). The spin entropy of rotating black hole have also been discussed (\cite{1979NCimB..51..262C}). If the black hole exists Cauchy horizon (CH) and event horizon (EH), we could discuss the products of these physical quantities, such as horizon and surface gravity etc. These properties are usually independent of the mass of the black hole, and depending on various charges, angular momenta and moduli. If these products are independent of the black hole mass, then these results may universal. The products of the thermodynamical features for black hole with rotational symmetry have been discussed e.g.,(\cite{2015arXiv150701547M, 2015JETPL.102..427P,2015MPLA...3050170P,2015PhLB..747...64P,2014EPJC...74.2887P,2011PhRvL.106l1301C,2013PhRvD..88d4014V}).

The phase transition of a black hole is a very interesting topics e.g., (\cite{1989CQGra...6.1909D,1978RPPh...41.1313D,1977RSPSA.353..499D,2015PhRvD..91h4009G,1982CMaPh..87..577H,2016PhLB..760.112, 2017ApJ..835.247,2008PhRvD..78b4016R,2011Ap&SS.333..449S,2014MPLA...2950057T,2012GReGr..44.2181T,2016arXiv160804176M}). For the Schwarzschild black hole, the heat capacity is always negative and no phase transition exists. But for Reissner¨CNordstr?m (RN) black hole, the phase transition would happen due to the existence of charge. If black hole has angular momentum and charge, the second order phase transition would happen. On the other hand, the geometry method can describe the black hole thermodynamics, and it is called the thermodynamic geometry (\cite{1975JChPh..63.2479W,1979PhRvA..20.1608R}).

If black hole is surrounded by quintessence or other  matters, the phase transition may be changeed, and this is a very interesting question. For quintessence matter around Schwarzschild black hole, the solution of Einstein$'$s field equation have been obtained by Kiselev (\cite{2003CQGra..20.1187K}). The solution is determined by the equation of state $\omega$ and strength parameter $\alpha$. Interestingly, the solution can also describe the radiation and dust around black holes (\cite{2015arXiv150804761M}). Recently the rotational Kiselev black hole (KBH) solution and charge Kiselev black hole solution have been obtained (\cite{2016EPJC...76..222G,2016Ap&SS.361..269O,2014PhRvD..90f4041A,2015arXiv151201498T,2016arXiv160902045X}). Following these works, the thermodynamical properties and phase transition have been studied in KBH for different situations e.g.,(\cite{2015arXiv150804761M,2015PhRvD..91f4049A,2017arXiv171008642J,2017GReGr..49...79M,2014MPLA...2950057T,2012GReGr..44.2181T,2018CoTPh..69..173W,2011ChPhL..28j0403W,2013ChPhB..22c0402W,2016arXiv161005454X,2017PhLB..764..100Z}). But for rotational KBH, the thermodynamic properties and phase transitions have not been studied. Compared with the  spherical symmetry KBH, the thermodynamics properties and phase transitions of rotational KBH may have new phenomenon. Therefore, it will be very interesting to study the thermodynamics and phase transition under rotational KBH.

In this work, we investigate the approximate solution of  thermodynamic quantities in rotational KBH, the products of thermodynamic quantities, the generalized Smarr formula and Christodoulou-Ruffini mass formula. We also calculate the phase transition and thermodynamic geometry for KBH in the case of radiation ($\omega=1/3$). The outline of this article are as follows. In Section \ref{properties}, we calculate the thermodynamic quantities and some properties for KBH. In Section \ref{phase-tr}, the phase transition and thermodynamic geometry have been discussed for radiation ($\omega=1/3$). The Summary is presented in the last Section.

\section{THERMODYNAMIC PROPERTIES IN ROTATIONAL KBH}
\label{properties}
The rotational symmetry solution for Einstein$'$s field equation with quintessence (or other energy matter) have been given by (\cite{2014PhRvD..90f4041A,2015arXiv151201498T})
\begin{equation}
ds^{2}=-(1-\dfrac{2Mr+\alpha r^{1-3\omega}}{\Sigma^{2}})dt^{2}+\dfrac{\Sigma^{2}}{\Delta_{r}}dr^{2}-\dfrac{2a \sin^{2}\theta(2Mr+\alpha r^{1-3\omega})}{\Sigma^{2}}d\phi dt+\Sigma^{2}d\theta^{2}$$$$
+\sin^{2}\theta (r^{2}+a^{2}+a^{2}\sin^{2}\theta\dfrac{2Mr+\alpha r^{1-3\omega}}{\Sigma^{2}})d\phi^{2},
\label{1}
\end{equation}
where
\begin{equation}
\Delta_{r}=r^{2}-2Mr+a^{2}-\alpha r^{1-3\omega},~~~~~\Sigma^{2}=r^{2}+a^{2}\cos^{2}\theta.
\label{2}
\end{equation}
In this black hole space-time metric, $M$ is the black hole mass, $a$ is the black hole spin, $\alpha$ is the strength parameter, the equation of state $\omega$ is define by $p=\omega\rho$, where $p$ is the pressure and $\rho$ is the energy density of quintessence. If the energy matter is quintessence, the equation of state $\omega$ will satisfy the $-1<\omega<-1/3$, the quintessence can lead to the universe acceleration. In the case of $\omega=1/3$, the energy matter represents radiation, and the case of $\omega=0$ represents dust around the Kerr black hole.

The black hole horizon is determined by the following equation
\begin{equation}
\Delta_{r}=r^{2}-2Mr+a^{2}-\alpha r^{1-3\omega}=0,
\label{3}
\end{equation}
this equation has three horizons if $-1<\omega<-1/3$, including the CH, EH and cosmological horizon. In the case of $\omega>-1/3$, the cosmological horizon will disappear and the eq. (\ref{3}) exists two horizons. Here we only consider the CH and EH, the cosmological horizon does not exist, but this situation is also interesting because the metric describes other energy matters such as radiation and dust around the Kerr black hole. In general $\alpha$ is very small. In the horizon equation (\ref{3}), we will use the perturbation method to calculate horizon radius (\cite{2011A, 2011B}). In this work, we make the quantities relating to $\alpha$ one-order, the horizon equation (\ref{3}) is written as
\begin{equation}
r_{\pm}=R_{\pm}+\epsilon_{\pm},
\label{4}
\end{equation}
where $R_{\pm}$ is the horizon radius of Kerr black hole. $r_{+}$ is the EH $\mathscr{H^{+}}$, $r_{-}$ is the CH $\mathscr{H^{-}}$. Substituting eq. (\ref{4}) into eq. (\ref{3}), we expand the equation for the first order term. The eq. (\ref{3}) becomes
\begin{equation}
(2R_{\pm}-2M-\alpha(1-3\omega)R^{-3\omega}_{\pm})\epsilon_{\pm}-\alpha R^{1-3\omega}_{\pm}=0,
\label{5}
\end{equation}
the solution of equation (\ref{5}) satisfies the condition of $\epsilon_{\pm}=0$ for $\alpha=0$, therefore the solution are given by
\begin{equation}
\epsilon_{\pm}=\dfrac{\alpha R^{1-3\omega}_{\pm}}{2R_{\pm}-2M-\alpha(1-3\omega)R^{-3\omega}_{\pm}}\simeq \pm\dfrac{\alpha(M\pm\sqrt{M^{2}-a^{2}})^{1-3\omega}}{2\sqrt{M^{2}-a^{2}}},~~~
R_{\pm}=M\pm\sqrt{M^{2}-a^{2}},
\label{6}
\end{equation}
where
\begin{equation}
r_{+}\simeq M+\sqrt{M^{2}-a^{2}}+\dfrac{\alpha (M+\sqrt{M^{2}-a^{2}})^{1-3\omega}}{2\sqrt{M^{2}-a^{2}}-\alpha(1-3\omega)(M+\sqrt{M^{2}-a^{2}})^{-3\omega}}$$$$
\simeq M+\sqrt{M^{2}-a^{2}}+\dfrac{\alpha (M+\sqrt{M^{2}-a^{2}})^{1-3\omega}}{2\sqrt{M^{2}-a^{2}}},
\label{7}
\end{equation}

\begin{equation}
r_{-}\simeq M-\sqrt{M^{2}-a^{2}}+\dfrac{\alpha (M-\sqrt{M^{2}-a^{2}})^{1-3\omega}}{-2\sqrt{M^{2}-a^{2}}-\alpha(1-3\omega)(M-\sqrt{M^{2}-a^{2}})^{-3\omega}}$$$$
\simeq M-\sqrt{M^{2}-a^{2}}-\dfrac{\alpha (M-\sqrt{M^{2}-a^{2}})^{1-3\omega}}{2\sqrt{M^{2}-a^{2}}}.
\label{8}
\end{equation}
Here we will consider three cases: quintessence matter $-2/3$, radiation $1/3$ and dust $0$. Using these cases, we discuss thermodynamic quantities and it$'$s products in detail. The product of horizon for any $\omega$ is given by
\begin{equation}
r_{+}r_{-}=a^{2}+\dfrac{\alpha a^{2}}{2\sqrt{M^{2}-a^{2}}}(R^{-3\omega}_{+}-(R^{-3\omega}_{+}).
\label{9}
\end{equation}

Case I: Quintessence matter ($\omega=-2/3$). For this case, there are no exact solutions for $\Delta_{r}=0$. Form eq.  (7) and eq. (8), we find that
\begin{equation}
r_{\pm}\simeq M\pm\sqrt{M^{2}-a^{2}}\pm\dfrac{\alpha (M\pm\sqrt{M^{2}-a^{2}})^{3}}{2\sqrt{M^{2}-a^{2}}},
\label{10}
\end{equation}
and the product of the horizon is given by
\begin{equation}
r_{+}r_{-}=a^{2}+2Ma^{2}\alpha,
\label{11}
\end{equation}
If we don$'$t consider the quintessence matter around Kerr black hole, then $\alpha=0$, the product of the horizon will degenerate into Kerr black hole case. Since this product is a function of the black hole mass $M$, then  this result is not universal. This is not consistent with Kerr black hole case.

Case II: Radiation ($\omega=1/3$). For this case, there are two exact solutions for $\Delta_{r}=0$. Form eq. (7) and eq. (8), we find that
\begin{equation}
r_{\pm}=M\pm\sqrt{M^{2}-a^{2}+\alpha},
\label{12}
\end{equation}
and the product of the horizon is given by
\begin{equation}
r_{+}r_{-}=a^{2}-\alpha.
\label{13}
\end{equation}
If we don$'$t consider the radiation around Kerr black hole, then $\alpha=0$, the product of the horizon will degenerate into Kerr black hole case. Since this product is independent of the mass of the black hole $M$, then this result is universal. But this product depends on strength parameter $\alpha$.

Case III: Dust ($\omega=0$). For this case,there are two exact solutions for $\Delta_{r}=0$. Form eq. (7) and eq. (8), we find that
\begin{equation}
r_{\pm}=M+\dfrac{\alpha}{2}\pm\sqrt{M^{2}-a^{2}+M\alpha+\dfrac{\alpha^{2}}{4}},
\label{14}
\end{equation}
and the product of the horizon is given by
\begin{equation}
r_{+}r_{-}=a^{2}.
\label{15}
\end{equation}
For this case, the dust around Kerr black hole doesn$'$t change the product. Of course, this result is universal. This would be exactly the same as the Kerr black hole.

Generally speaking, for rotational KBH, the properties of horizon product  depend entirely on the equation of state $\omega$ and strength parameter $\alpha$. This is consistent with the physical characteristics of rotational KBH.

And then we will calculate the other thermodynamic quantities and the corresponding products. For rotational KBH, the horizon areas are given by
\begin{equation}
A_{\pm}=\int\sqrt{g}d\theta d\phi=\int^{2\pi}_{0}\int^{\pi}_{0}\sqrt{g_{\theta\theta}g_{\phi\phi}}d\theta d\phi=4\pi( r^{2}_{\pm}+a^{2})\simeq 4\pi(R^{2}_{\pm}+2R_{\pm}\epsilon_{\pm}+a^{2})$$$$=4\pi(2M^{2}\pm2M\sqrt{M^{2}-a^{2}})\pm
4\pi\dfrac{\alpha(M\pm\sqrt{M^{2}-a^{2}})^{2-3\omega}}{\sqrt{M^{2}-a^{2}}}.
\label{16}
\end{equation}
The semi-classical Bekenstein-Hawking entropy at $\mathscr{H^{\pm}}$ are given by (\cite{1971PhRvL..26.1344H})
\begin{equation}
S_{\pm}=\dfrac{A_{\pm}}{4}\simeq \pi(2M^{2}\pm2M\sqrt{M^{2}-a^{2}})\pm \pi\dfrac{\alpha(M\pm\sqrt{M^{2}-a^{2}})^{2-3\omega}}{\sqrt{M^{2}-a^{2}}}.
\label{17}
\end{equation}
The surface gravity of the rotational KBH are defined by
\begin{equation}
\kappa_{\pm}=\lim\limits_{r\rightarrow{r_{+}}}(-\dfrac{1}{2}\sqrt{\dfrac{g^{rr}}{-g^{'}_{tt}}g^{'}_{tt,r}})=\dfrac{r_{+}-r_{-}}{2(r^{2}_{\pm}+a^{2})}=\dfrac{R_{+}-R_{-}+\epsilon_{+}-\epsilon_{-}}{2((R_{\pm}+\epsilon_{\pm})^{2}+a^{2})}$$$$
\simeq\dfrac{\sqrt{M^{2}-a^{2}}}{2MR_{\pm}}+\dfrac{\alpha}{4MR_{\pm}\sqrt{M^{2}-a^{2}}}[\dfrac{1}{2}(R^{1-3\omega}_{+}+R^{1-3\omega}_{-})\mp\dfrac{\sqrt{M^{2}-a^{2}}R^{1-3\omega}_{\pm}}{M}],
\label{18}
\end{equation}
where $g^{'}_{tt}=g_{tt}-g^{2}_{t\phi}/g_{\phi\phi}$.

From the black hole thermodynamics, the Hawking temperature for $\mathscr{H^{\pm}}$ reads
\begin{equation}
T_{\pm}=\dfrac{\kappa_{\pm}}{2\pi}=\dfrac{R_{+}-R_{-}+\epsilon_{+}-\epsilon_{-}}{4\pi((R_{\pm}+\epsilon_{\pm})^{2}+a^{2})}$$$$
\simeq\dfrac{\sqrt{M^{2}-a^{2}}}{4\pi MR_{\pm}}+\dfrac{\alpha}{8\pi MR_{\pm}\sqrt{M^{2}-a^{2}}}[\dfrac{1}{2}(R^{1-3\omega}_{+}+R^{1-3\omega}_{-})\mp\dfrac{\sqrt{M^{2}-a^{2}}R^{1-3\omega}_{\pm}}{M}].
\label{19}
\end{equation}

For rotational KBH, the rotation velocity of the horizon reads
\begin{equation}
\Omega_{\pm}=\lim\limits_{r\rightarrow{r_{\pm}}}(\dfrac{-g_{t\phi}}{g_{\phi\phi}})=\dfrac{a}{r^{2}_{\pm}+a^{2}}.
\label{20}
\end{equation}

From eq. (16)-(20), one can find that horizon areas, Bekenstein-Hawking entropy, surface gravity and Hawking temperature are depends on Kerr black hole parameters $(M,a)$, the equation of state $\omega$   and strength parameter $\alpha$. But the rotation velocity of the horizon are independent of the equation of state $\omega$ and strength parameter $\alpha$.

By the expression for horizon areas $A_{\pm}$, Bekenstein-Hawking entropy $S_{\pm}$, surface gravity $\kappa_{\pm}$ and Hawking temperature $T_{\pm}$, we can calculate the Komar energy (\cite{1959PhRv..113..934K}), the product of Komar energy, surface gravity, Hawking temperature, areas and Bekenstein-Hawking entropy at $\mathscr{H^{\pm}}$, the results are as follows: The Komar energy is given by
\begin{equation}
E_{\pm}=2S_{\pm}T_{\pm}=\dfrac{(R^{2}_{\pm}+2R_{\pm}\epsilon_{\pm})(R_{+}-R_{-}+\epsilon_{+}-\epsilon_{-})}{2((R_{\pm}+\epsilon_{\pm})^{2}+a^{2})}$$$$
\simeq\dfrac{R_{\pm}\sqrt{M^{2}-a^{2}}}{2M}+\dfrac{R_{\pm}\alpha}{2M}[\dfrac{1}{2\sqrt{M^{2}-a^{2}}}(\dfrac{1}{2}(R^{1-3\omega}_{+}+R^{1-3\omega}_{-})\mp\dfrac{\sqrt{M^{2}-a^{2}}R^{1-3\omega}_{\pm}}{M})\pm R^{1-3\omega}_{\pm}].
\label{21}
\end{equation}
The product of surface gravities namely Hawking temperatures, reads
\begin{equation}
\kappa_{+}\kappa_{-}=4\pi^{2}T_{+}T_{-}=\dfrac{(R_{+}-R_{-}+\epsilon_{+}-\epsilon_{-})^{2}}{2((R_{+}+\epsilon_{+})^{2}+a^{2})(R_{-}+\epsilon_{-})^{2}+a^{2}))}$$$$
\simeq\dfrac{M^{2}-a^{2}}{4M^{2}a^{2}}+\dfrac{\alpha}{8M^{2}a^{2}}[R^{1-3\omega}_{+}+R^{1-3\omega}_{-}+\dfrac{\sqrt{M^{2}-a^{2}}}{M}(R^{1-3\omega}_{-}-R^{1-3\omega}_{+})].
\label{22}
\end{equation}
The product of Komar energies is
\begin{equation}
E_{+}E_{-}=\dfrac{(R^{2}_{+}+2R_{+}\epsilon_{+})(R^{2}_{-}+2R_{-}\epsilon_{-})(R_{+}-R_{-}+\epsilon_{+}-\epsilon_{-})}{2((R_{+}+\epsilon_{+})^{2}+a^{2})(R_{-}+\epsilon_{-})^{2}+a^{2}))}$$$$
\simeq\dfrac{a^{2}(M^{2}-a^{2})}{4M^{2}}+\dfrac{a^{2}\alpha\sqrt{M^{2}-a^{2}}}{4M^{2}}[\dfrac{1}{2\sqrt{M^{2}-a^{2}}}(R^{1-3\omega}_{+}+R^{1-3\omega}_{-})+(1-\dfrac{1}{2M})(R^{1-3\omega}_{+}-R^{1-3\omega}_{-})].
\label{23}
\end{equation}
The product of horizon areas, namely the Bekenstein-Hawking entropies, reads
\begin{equation}
A_{+}A_{-}=16S_{+}S_{-}=16\pi^{2}(R^{2}_{+}+2R_{+}\epsilon_{+})(R^{2}_{-}+2R_{-}\epsilon_{-})
\simeq a^{4}+\dfrac{\alpha a^{4}}{\sqrt{M^{2}-a^{2}}}(R^{-3\omega}_{+}-R^{-3\omega}_{-}).
\label{24}
\end{equation}
From the eq.(22)-(24), one can find that the products of surface gravities, Hawking temperatures, Komar energies, horizon areas and Bekenstein-Hawking entropies are dependent of the Kerr black hole parameters $(M,a)$, the equation of state $\omega$ and strength parameter $\alpha$, especially for the equation of state $\omega$. Then, respectively for the quintessence matter, radiation and dust, we discuss these products in details.

Case I: Quintessence matter ($\omega=-2/3$). For this case, the expression for these products (eq. (22)-(24)) becomes
\begin{equation}
\kappa_{+}\kappa_{-}=4\pi^{2}T_{+}T_{-}\simeq\dfrac{M^{2}-a^{2}}{4M^{2}a^{2}}+\alpha(\dfrac{1}{2M}-\dfrac{a^{2}}{4M^{3}}),
\label{25}
\end{equation}

\begin{equation}
E_{+}E_{-}\simeq\dfrac{a^{2}(M^{2}-a^{2})}{4M^{2}}+\dfrac{a^{2}\alpha\sqrt{M^{2}-a^{2}}}{4M^{2}}((4M^{2}-a^{2})(4M^{3}-4Ma^{2}+2a^{2})-4M^{2}a^{2}),
\label{26}
\end{equation}

\begin{equation}
A_{+}A_{-}\simeq 16S_{+}S_{-}=a^{4}+4Ma^{4}\alpha.
\label{27}
\end{equation}
If one does not consider the quintessence matter around Kerr black hole, then $\alpha=0$, these products (eq. (25)-(27)) will degenerate into Kerr black hole case. Since these products are function of the black hole mass $M$, then these results are not universal.

Case II: Radiation ($\omega=1/3$). For this case, the expression for these products (eq. (22)-(24)) becomes
\begin{equation}
\kappa_{+}\kappa_{-}=4\pi^{2}T_{+}T_{-}=\dfrac{M^{2}-a^{2}}{4M^{2}a^{2}}+\dfrac{\alpha}{4M^{2}a^{2}},
\label{28}
\end{equation}

\begin{equation}
E_{+}E_{-}=\dfrac{a^{2}(M^{2}-a^{2})}{4M^{2}}+\dfrac{a^{2}\alpha}{4M^{2}},
\label{29}
\end{equation}

\begin{equation}
A_{+}A_{-}=16S_{+}S_{-}=a^{4}-2Ma^{2}\alpha.
\label{30}
\end{equation}
If one does not consider the radiation around Kerr black hole, then $\alpha=0$, these products (eq. (28)-(30)) will degenerate into Kerr black hole case. Since these products are function of the black hole mass $M$, then these results are not universal.

Case III: Dust ($\omega=0$). For this case, the expression for these products (eq. (22)-(24)) becomes
\begin{equation}
\kappa_{+}\kappa_{-}=4\pi^{2}T_{+}T_{-}=\dfrac{M^{2}-a^{2}}{4M^{2}a^{2}}+\dfrac{\alpha}{4M^{3}},
\label{31}
\end{equation}

\begin{equation}
E_{+}E_{-}=\dfrac{a^{2}(M^{2}-a^{2})}{4M^{2}}+\dfrac{a^{2}\alpha}{4M^{3}}(2M(M^{2}-a^{2})+a^{2}),
\label{32}
\end{equation}

\begin{equation}
A_{+}A_{-}=16S_{+}S_{-}=a^{4}.
\label{33}
\end{equation}
If one does not consider the radiation around Kerr black hole, then $\alpha=0$, these products (eq. (28)-(30)) will degenerate into Kerr black hole case. For products of surface gravity, Hawking temperature and Komar energy, these products are function of the black hole mass $M$, then these results are not universal. On the other hand. For products of horizon areas and Bekenstein-Hawking entropy, the dust around Kerr black hole doesn$'$t change these products, then these results are universal.

\subsection{Generalized Smarr Formula}
\label{Smarr}
From the eq. (\ref{3}),  for the general equation of state $\omega$, the general expression (\cite{1973PhRvD...7..289S, 1973PhRvL..30...71S}) connecting Bekenstein-Hawking entropy, black hole angular momentum and black hole mass is
\begin{equation}
M=\dfrac{1}{2}\sqrt{\dfrac{S}{\pi}-\dfrac{J^{2}}{M^{2}}}+\dfrac{J^{2}}{2M^{2}}\sqrt{\dfrac{\pi M^{2}}{SM^{2}-\pi J^{2}}}-\dfrac{\alpha}{2}(\dfrac{S}{\pi}-\dfrac{J^{2}}{M^{2}})^{-\dfrac{3\omega}{2}}.
\label{44}
\end{equation}
Generally speaking, one cannot have an analytic expression for black hole mass. For radiation ($\omega=1/3$) around Kerr black hole case, we can get an analytic expression for mass $M$. For this case, the eq. (\ref{44}) will turn into
\begin{equation}
S=\pi(2M^{2}+2\sqrt{M^{4}-J^{2}+M^{2}\alpha}+\alpha).
\label{47}
\end{equation}
Therefore, we obtain the $M=M(S,J,\alpha)$, which reads
\begin{equation}
M^{2}=\dfrac{S}{4\pi}-\dfrac{\alpha}{2}+\dfrac{\pi\alpha^{2}}{4S}+\dfrac{\pi J^{2}}{S}.
\label{48}
\end{equation}
One can define the quantities related to Bekenstein-Hawking entropy $S$, angular momentum $J$ and strength parameter $\alpha$, such as the effective surface tension and angular velocity are given by
\begin{equation}
T=(\dfrac{\partial M}{\partial S})_{J}=\dfrac{\dfrac{1}{2\pi}-\dfrac{\pi\alpha^{2}}{8S^{2}}-\dfrac{\pi J^{2}}{2S^{2}}}{\sqrt{\dfrac{S}{4\pi}-\dfrac{\alpha}{2}+\dfrac{\pi\alpha^{2}}{4S}+\dfrac{\pi J^{2}}{S}}},
\label{49}
\end{equation}

\begin{equation}
\Omega=(\dfrac{\partial M}{\partial J})_{S}=\dfrac{\pi}{M}\dfrac{J}{S}.
\label{50}
\end{equation}

According to the first law of thermodynamics, the mass variation of black hole is determined by the horizon areas and angular momentum as
\begin{equation}
dM=TdA+\Omega dJ,
\label{62}
\end{equation}
where $T$ is the effective surface tension. Its expression is
\begin{equation}
T_{\pm}=\dfrac{1}{M}(\dfrac{1}{2\pi}-\dfrac{\pi\alpha^{2}}{8S^{2}}-\dfrac{\pi J^{2}}{2S^{2}})
=\dfrac{1}{M}(\dfrac{1}{2\pi}-(\dfrac{\pi\alpha^{2}}{8}-\dfrac{\pi M^{2}a^{2}}{2})\dfrac{1}{\pi^{2}(2M^{2}\pm 2M\sqrt{M^{2}-a^{2}+\alpha}+\alpha)^{2}})  $$$$
=\dfrac{1}{8\pi}\dfrac{\sqrt{M^{2}-a^{2}+\alpha}}{2M^{2}\pm 2M\sqrt{M^{2}-a^{2}+\alpha}+\alpha}=\dfrac{\kappa_{\pm}}{8\pi},
\label{63}
\end{equation}
where $S=\pi(2M^{2}\pm 2M\sqrt{M^{2}-a^{2}+\alpha}+\alpha)$ and $J=Ma$.

The angular velocity $\Omega$ is given by
\begin{equation}
\Omega=\dfrac{\pi J}{MS}
=\dfrac{\pi Ma}{M\pi (2M^{2}\pm 2(M^{4}-J^{2})^{1/2})}
=\dfrac{a}{2Mr_{\pm}}
=\dfrac{a}{r^{2}_{\pm}+a^{2}}
=\Omega_{\pm}.
\label{64}
\end{equation}

From the above analysis, we found that when consider radiation around Kerr black hole, the black hole mechanics quantity  satisfy the first law of thermodynamics. This is consistent with the case of the Kerr black hole.

\subsection{Christodoulou-Ruffini Mass Formula}
\label{crmass}
The black hole mass can decrease or increase, but the irreducible mass $M_{irr}$ of black hole can not decrease (\cite{1970PhRvL...1596,1971PhRvD...4.3552C}). The increase of $M_{irr}$ will lead to many physical processes in black hole physics. According to the second law of thermodynamics for black holes. The areas of horizon or Bekenstein-Hawking entropy all increase, e.g., $dS=dA/4\geq 0$, which lead a relation between horizon area and irreducible mass for black hole.

In rotational KBH, the black hole has EH and CH, thus the irreducible mass and horizon areas satisfy the following equation

\begin{equation}
M_{irr\pm}=\sqrt{\dfrac{A_{\pm}}{16\pi}}=\sqrt{\dfrac{r^{2}_{\pm}+a^{2}}{4}}
=(\dfrac{1}{2}M^{2}\pm\dfrac{1}{2}M\sqrt{M^{2}-a^{2}}\pm
\dfrac{\alpha(M\pm\sqrt{M^{2}-a^{2}})^{2-3\omega}}{4\sqrt{M^{2}-a^{2}}})^{1/2},
\label{100}
\end{equation}
where $M_{irr+}$ and $M_{irr-}$ are black hole irreducible masses on the EH and CH. The product of the irreducible masses is given by
\begin{equation}
M_{irr+}M_{irr-}=\dfrac{\sqrt{A_{+}A_{-}}}{16\pi}
=\dfrac{\sqrt{a^{4}+\dfrac{\alpha a^{4}}{\sqrt{M^{2}-a^{2}}}(R^{-3\omega}_{+}-R^{-3\omega}_{-})}}{16\pi}.
\label{101}
\end{equation}
The rest mass of rotating black hole is given by the following expression
\begin{equation}
M^{2}=M^{2}_{irr\pm}+\dfrac{J^{2}}{4M^{2}_{irr\pm}}
\label{102}
\end{equation}

From eq. (\ref{100}-\ref{102}), we found that the irreducible massed and it$'$s product are depends on the equation of state $\omega$ and strength parameter $\alpha$. Its physical properties are quite different from Kerr black holes.

\section{PHASE TRANSITION AND THERMODYNAMIC GEOMETRY}
\label{phase-tr}
In this section, we study the phase transition in rotational KBH.
The heat capacity of the black hole is an interesting thermodynamical property presenting the stability and instability of the black hole. General black hole has a negative heat capacity to be unstable and produce Hawking radiation (\cite{1982PhRvD..25..330G}). But if the black hole brings angular momentum or charge, the heat capacity can be positive, and the phase transition will occur. (\cite{1977RSPSA.353..499D}).

For the KBH (spherically symmetric), the phase transition has been discussed by many authers e.g., (\cite{2015arXiv150804761M,2015PhRvD..91f4049A,2017arXiv171008642J,2017GReGr..49...79M,2018CoTPh..69..173W,2011ChPhL..28j0403W,2016arXiv161005454X,2017PhLB..764..100Z,2014MPLA...2950057T,2012GReGr..44.2181T,2011A,2016Ap&SS.361..161G,2013ChPhB..22c0402W}). For rotational KBH, we want to analyze the phase transition (radiation case). We write the black hole mass as a function of Bekenstein-Hawking entropy and angular momentum $M=M(S,J,\alpha,\omega)$ is given by
\begin{equation}
M=\dfrac{1}{2}\sqrt{\dfrac{S}{\pi}-\dfrac{J^{2}}{M^{2}}}+\dfrac{J^{2}}{2M^{2}}\sqrt{\dfrac{\pi M^{2}}{SM^{2}-\pi J^{2}}}-\dfrac{\alpha}{2}(\dfrac{S}{\pi}-\dfrac{J^{2}}{M^{2}})^{-\dfrac{3\omega}{2}}.
\label{130}
\end{equation}

From the subsection \ref{Smarr}, we know that the first law of thermodynamics for KBH is given by
\begin{equation}
dM=TdS+\Omega dJ,
\label{131}
\end{equation}
From the eq. (\ref{130}), we obtain Hawking temperature and angular velocity reads
\begin{equation}
T=(\dfrac{\partial M}{\partial S})_{J}=\dfrac{\dfrac{1}{4\pi}+\dfrac{3\alpha}{4\pi}\omega(\dfrac{S}{\pi}-\dfrac{J^{2}}{M^{2}})^{-\dfrac{3}{2}\omega-\dfrac{1}{2}}}{\sqrt{\dfrac{S}{\pi}-\dfrac{J^{2}}{M^{2}}}-\dfrac{J^{2}}{2M^{3}}+\dfrac{\pi J^{2}(4M+J^{2})}{4M^{2}(SM^{2}-\pi J^{2})}-\dfrac{3\alpha J^{2}}{2M^{3}}\omega(\dfrac{S}{\pi}-\dfrac{J^{2}}{M^{2}})^{-\dfrac{3}{2}\omega-\dfrac{1}{2}}},
\label{132}
\end{equation}

\begin{equation}
\Omega=(\dfrac{\partial M}{\partial J})_{S}=$$$$
\dfrac{-\dfrac{J}{2M^{2}}+\dfrac{\pi J}{SM^{2}-\pi J^{2}}+\dfrac{\pi^{2}J^{3}}{2(SM^{2}-\pi J^{2})^{2}}-\dfrac{3\alpha}{2}\omega\dfrac{J}{M^{2}}(\dfrac{S}{\pi}-\dfrac{J^{2}}{M^{2}})^{-\dfrac{3}{2}\omega-\dfrac{1}{2}}}{\sqrt{\dfrac{S}{\pi}-\dfrac{J^{2}}{M^{2}}}-\dfrac{J^{2}}{2M^{3}}+\dfrac{\pi J^{2}}{M(SM^{2}-\pi J^{2})}+\dfrac{\pi^{2}J^{4}}{2M(SM^{2}-\pi J^{2})^{2}}-\dfrac{3\alpha\omega J^{2}}{2M^{3}}(\dfrac{S}{\pi}-\dfrac{J^{2}}{M^{2}})^{-\dfrac{3}{2}\omega-\dfrac{1}{2}}}.
\label{133}
\end{equation}

To simplify these equations, we set
\begin{equation}
T=\dfrac{X}{Y},
\label{134}
\end{equation}
where $X$ and $Y$ are
\begin{equation}
X=\dfrac{1}{4\pi}+\dfrac{3\alpha}{4\pi}\omega(\dfrac{S}{\pi}-\dfrac{J^{2}}{M^{2}})^{-\dfrac{3}{2}\omega-\dfrac{1}{2}},$$$$
Y=\sqrt{\dfrac{S}{\pi}-\dfrac{J^{2}}{M^{2}}}-\dfrac{J^{2}}{2M^{3}}+\dfrac{\pi J^{2}(4M+J^{2})}{4M^{2}(SM^{2}-\pi J^{2})}-\dfrac{3\alpha J^{2}}{2M^{3}}\omega(\dfrac{S}{\pi}-\dfrac{J^{2}}{M^{2}})^{-\dfrac{3}{2}\omega-\dfrac{1}{2}}.
\label{300}
\end{equation}
The heat capacity $C_{J}$ can be express as
\begin{equation}
C_{J}=T(\dfrac{\partial S}{\partial T})_{J}=\dfrac{XY}{\dfrac{\partial X}{\partial S}Y-X\dfrac{\partial Y}{\partial S}},
\label{135}
\end{equation}
where $\dfrac{\partial X}{\partial S}Y-X\dfrac{\partial Y}{\partial S}$ is given by
\begin{equation}
\dfrac{\partial X}{\partial S}Y-X\dfrac{\partial Y}{\partial S}=-\dfrac{3\alpha\omega}{8\pi}(3\omega+1)(\dfrac{S}{\pi}-\dfrac{J^{2}}{M^{2}})^{-\dfrac{3}{2}(\omega+1)}(\dfrac{1}{\pi}+\dfrac{2J^{2}}{M^{3}}T)-\dfrac{1}{4\pi}(1+3\alpha\omega(\dfrac{S}{\pi}-$$$$\dfrac{J^{2}}{M^{2}})^{-\dfrac{1}{2}(3\omega+1)})[\dfrac{\dfrac{1}{\pi}+\dfrac{2J^{2}}{M^{3}}T}{2\sqrt{\dfrac{S}{\pi}-\dfrac{J^{2}}{M^{2}}}}+\dfrac{3J^{2}}{2M^{4}}T-
\dfrac{\pi J^{2}(4M+J^{2})}{4(SM^{2}-\pi J^{2})^{2}}-$$$$\dfrac{2\pi J^{2}((SM^{2}-\pi J^{2})(2M^{2}+MJ^{2})+SM^{3}(4M+J^{2}))T}{4M^{4}(SM^{2}-\pi J^{2})^{2}}+\dfrac{3\alpha J^{2}\omega}{4M^{3}}(3\omega+1)(\dfrac{S}{\pi}-\dfrac{J^{2}}{M^{2}})^{-\dfrac{3}{2}(\omega+1)}$$$$
(\dfrac{1}{\pi}+\dfrac{2J^{2}}{M^{3}}T)+\dfrac{9\alpha J^{2}}{2M^{4}}\omega(\dfrac{S}{\pi}-\dfrac{J^{2}}{M^{2}})^{-\dfrac{3}{2}\omega-\dfrac{1}{2}}T].
\label{136}
\end{equation}
The phase transition will occur when the following condition is  satisfy
\begin{equation}
\dfrac{\partial X}{\partial S}Y=X\dfrac{\partial Y}{\partial S}.
\label{137}
\end{equation}

For general equation of state $\omega$, we can not get an expression for heat capacity $C_{J}$. Here we take radiation case ($\omega=1/3$) to study the feature of phase transition. The generalized Smarr mass function has been obtained in subsection \ref{Smarr} as
\begin{equation}
M^{2}=\dfrac{S}{4\pi}-\dfrac{\alpha}{2}+\dfrac{\pi\alpha^{2}}{4S}+\dfrac{\pi J^{2}}{S}.
\label{138}
\end{equation}
Through this equation, we obtain the expression for $T, \Omega, C_{J}$ as
\begin{equation}
T=\dfrac{\dfrac{1}{4\pi}-\dfrac{\pi\alpha^{2}}{4S^{2}}-\dfrac{\pi J^{2}}{S^{2}}}{2\sqrt{\dfrac{S}{4\pi}-\dfrac{\alpha}{2}+\dfrac{\pi\alpha^{2}}{4S}+\dfrac{\pi J^{2}}{S}}},
\label{139}
\end{equation}

\begin{equation}
\Omega=\dfrac{\pi J}{MS}=\dfrac{\pi J}{S\sqrt{\dfrac{S}{4\pi}-\dfrac{\alpha}{2}+\dfrac{\pi\alpha^{2}}{4S}+\dfrac{\pi J^{2}}{S}}},
\label{140}
\end{equation}

\begin{equation}
C_{J}=\dfrac{2(\dfrac{S}{4\pi}-\dfrac{\alpha}{2}+\dfrac{\pi\alpha^{2}}{4S}+\dfrac{\pi J^{2}}{S})(\dfrac{1}{4\pi}-\dfrac{\pi\alpha^{2}}{4S^{2}}-\dfrac{\pi J^{2}}{S^{2}})}{(\dfrac{\pi\alpha^{2}}{S^{3}}+\dfrac{4\pi J^{2}}{S^{3}})(\dfrac{S}{4\pi}-\dfrac{\alpha}{2}+\dfrac{\pi\alpha^{2}}{4S}+\dfrac{\pi J^{2}}{S})-(\dfrac{1}{4\pi}-\dfrac{\pi\alpha^{2}}{4S^{2}}-\dfrac{\pi J^{2}}{S^{2}})^{2}}.
\label{141}
\end{equation}
The condition of phase transition is
\begin{equation}
(\dfrac{\pi\alpha^{2}}{S^{3}}+\dfrac{4\pi J^{2}}{S^{3}})(\dfrac{S}{4\pi}-\dfrac{\alpha}{2}+\dfrac{\pi\alpha^{2}}{4S}+\dfrac{\pi J^{2}}{S})=(\dfrac{1}{4\pi}-\dfrac{\pi\alpha^{2}}{4S^{2}}-\dfrac{\pi J^{2}}{S^{2}})^{2}.
\label{142}
\end{equation}

In order to analyze the critical point of the phase transition, we set $M=1$ to obtain $J=a$ and $S=\pi(2+\alpha+2\sqrt{1+\alpha-a^{2}})$. The eq. (\ref{142}) becomes
\begin{equation}
(2+\alpha+2\sqrt{1+\alpha-a^{2}})(\alpha^{2}+4a^{2})=(a^{2}+\dfrac{\alpha^{2}}{4}-\dfrac{1}{4}(8+8\alpha+\alpha^{2}-4a^{2}+4(2+\alpha)\sqrt{1+\alpha-a^{2}}))^{2}.
\label{143}
\end{equation}

\begin{table*}
\renewcommand{\arraystretch}{1.0}
\caption[]{At the critical point, variation of the strength parameter $\alpha$ versus black hole spin $a$. Where black hole mass $M=1$ and the equation of state $\omega=1/3$.}
\begin{center}
\begin{tabular}{lcccccccc}
\hline
\hline
 parameter$\alpha$     &0.0000 & 0.0010 & 0.0100 & 0.2000 & 0.5000  & 0.8000 & 0.9500 & 1\\
\hline
 black hole spin $a$   &0.6813 & 0.6817 & 0.6855 & 0.7601 & 0.8612 & 0.9480 & 0.9873 & 1\\
\hline
\end{tabular}
\end{center}
\label{tab:ks}
\end{table*}

Through numerical calculation, we find that the phase transition with $a\leq 1$ demands $\alpha\leq 1$ (of course, this condition is always satisfying). For different strength parameter $\alpha$, the critical black hole spin $a$ is given in Table 1. We find that the existence of radiation will boost the critical spin $a$ of black hole. When strength parameter $\alpha=0$, it reduces to Kerr black hole case (\cite{1977RSPSA.353..499D}). When black hole spin  $a=0$, it reduces to KBH and the phase transition is also second order phase transition e.g., (\cite{2016Ap&SS.361..161G,2015arXiv150804761M,2014MPLA...2950057T}). Through calculating the derivative of Gibbs free energy, we find that the phase transition is a second order phase transition for the rotational KBH in radiation case.

\begin{figure}[htbp]
  \centering
  \includegraphics[scale=0.55]{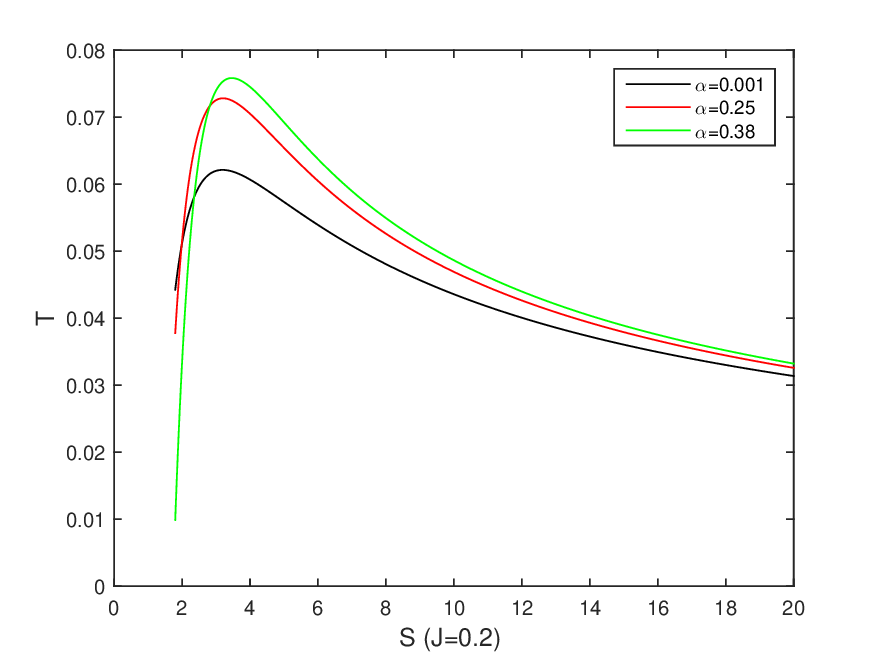}
  \includegraphics[scale=0.55]{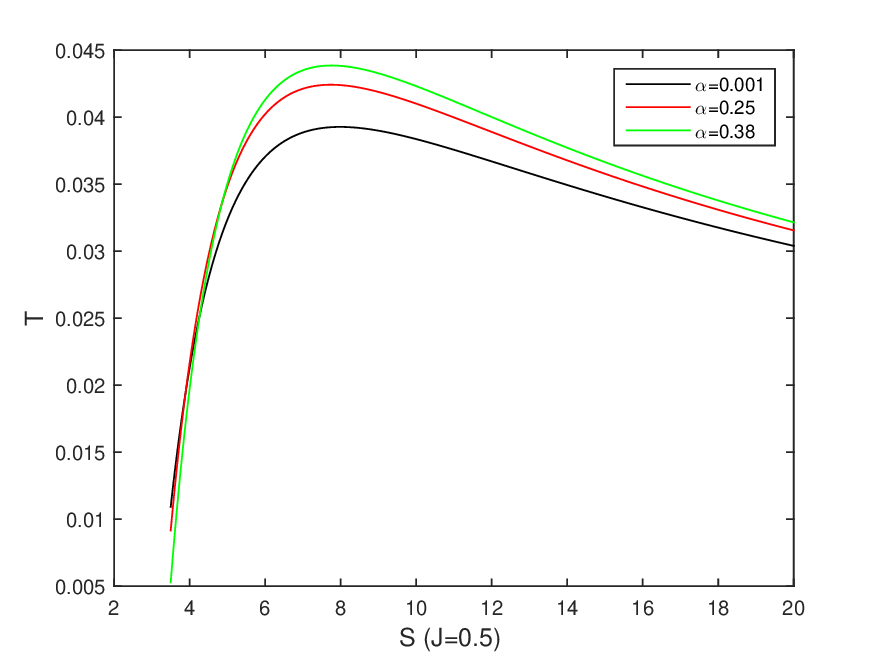}
  \caption{Hawking temperature $T$ versus Bekenstein-Hawking entropy $S$ for different strength parameter $\alpha$. Here we consider the radiation case $\omega=1/3$.}
  \label{fig:1}
\end{figure}

\begin{figure}[htbp]
  \centering
  \includegraphics[scale=0.55]{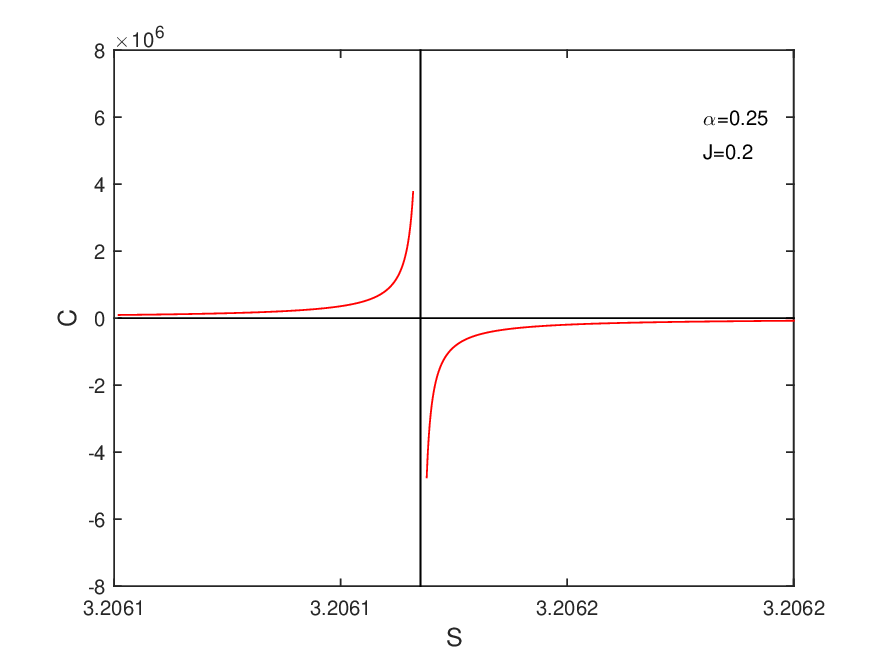}
  \includegraphics[scale=0.55]{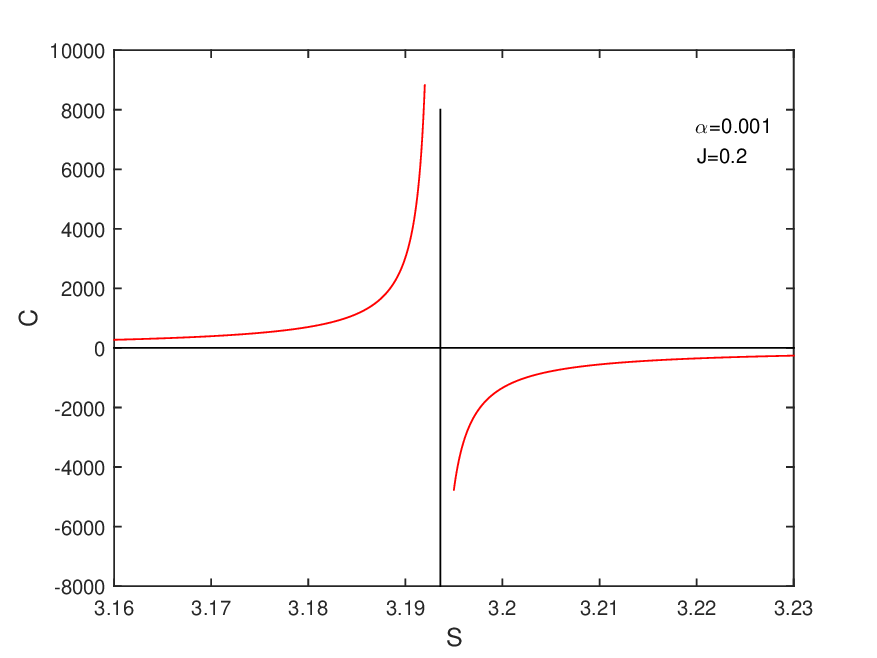}
  \caption{Variation of the heat capacity $C_{J}$ versus Bekenstein-Hawking entropy $S$ at constant angular momentum for different strength parameter $\alpha$. Here we consider the radiation case $\omega=1/3$.}
  \label{fig:2}
\end{figure}

We plot the variation of the Hawking temperature $T$ versus the Bekenstein-Hawking entropy $S$ for different strength parameter $\alpha$ and angular momentum $J$ in Fig.\ref{fig:1}. The variation of the heat capacity $C_{J}$ versus the Bekenstein-Hawking entropy $S$ for different strength parameter $\alpha$ is shown in Fig.\ref{fig:2}. From Fig.\ref{fig:1}, we find that the Hawking temperature will decrease when strength parameter $\alpha$ decreases. From Fig.\ref{fig:2}, we find that the critical point shifts to higher Bekenstein-Hawking entropy when the strength parameter $\alpha$ increases.

Following we discuss thermodynamic geometry of rotational KBH. For radiation situation $\omega=1/3$, the two horizon can be obtained in eq. (\ref{12}). The horizon equation $\Delta=r^{2}-2Mr+a^{2}-\alpha=0$ can lead to the relation

\begin{equation}
r_{+}+r_{-}=2M,~~~~~J=\dfrac{1}{2}(r_{+}+r_{-})\sqrt{\alpha+r_{+}r_{-}}.
\label{160}
\end{equation}

In black hole thermodynamics, the Bekenstein-Hawking entropy of the rotational KBH is given by
\begin{equation}
S=\dfrac{1}{4}A=\pi(r^{2}_{+}+a^{2})=\pi(r^{2}_{+}+\alpha+r_{+}r_{-}).
\label{161}
\end{equation}

According to the first law of black hole thermodynamics, the Hawking temperature and angular velocity are
\begin{equation}
T=(\dfrac{\partial M}{\partial S})_{J}=\dfrac{r^{4}_{+}+(r^{2}_{+}-2r_{+}r_{-}-r^{2}_{-})(\alpha+r_{+}r_{-})+(\alpha+r_{+}r_{-})^{2}-\alpha^{2}}{4\pi(r_{+}+r_{-})(r^{2}_{+}+\alpha+r_{+}r_{-})^{2}}
\label{162}
\end{equation}
and
\begin{equation}
\Omega=(\dfrac{\partial M}{\partial J})_{S}=\dfrac{\sqrt{\alpha+r_{+}r_{-}}}{r^{2}_{+}+\alpha+r_{+}r_{-}}.
\label{163}
\end{equation}
Similar to the Kerr black hole case, By correspondence $(\Omega,J)\longrightarrow (V,P)$, we can establish thermodynamics for rotational KBH. The internal energy of the rotational KBH should include the kinetic energy of rotation as
\begin{equation}
U=M-\Omega J=\dfrac{(r_{+}+r_{-})r^{2}_{+}}{2(r^{2}_{+}+\alpha+r_{+}r_{-})},
\label{164}
\end{equation}
then the first law of thermodynamics could expression as
\begin{equation}
dU=TdS-Jd\Omega.
\label{165}
\end{equation}
Through defining thermodynamics metric by following formalism e.g., (\cite{1975JChPh..63.2479W,1979PhRvA..20.1608R})
\begin{equation}
g_{\mu\nu}=\dfrac{\partial^{2}}{\partial x^{\mu}\partial x^{\nu}}S(U,\Omega),
\label{166}
\end{equation}
where $\mu,\nu=1,2$ and $x^{1}=U,x^{2}=\Omega$, we calculate Ruppeiner metric as
\begin{equation}
g_{\mu\nu}={\left(\begin{array}{cc}
\dfrac{-1}{T^{2}}(\dfrac{\partial T}{\partial U})&\dfrac{-1}{T^{2}}(\dfrac{\partial T}{\partial \Omega})\\
\dfrac{-1}{T^{2}}(\dfrac{\partial T}{\partial \Omega})&\dfrac{1}{T}\dfrac{\partial J}{\partial \Omega}-\dfrac{J}{T^{2}}\dfrac{\partial J}{\partial \Omega}
\end{array}
\right)}.
\label{167}
\end{equation}
For this metric, we can calculate the scalar curvature of rotational KBH, the results are as follows
\begin{equation}
R=g_{\mu\nu}R^{\mu\nu}=H(r_{+},r_{-})/((r_{+}-r_{-})([\alpha^{2}+(r_{+}+r_{-})^{2}(\alpha+r_{+}r_{-})][(r_{+}+r_{-})^{2}(6r^{2}_{+}+3\alpha+3r_{+}r_{-})+$$$$
\alpha(r_{+}+r_{-})r_{+}+\alpha^{2}]-(r^{2}_{+}+\alpha+r_{+}r_{-})^{4}-2r^{2}_{-})),
\label{168}
\end{equation}
where $H$ are the function of EH $r_{+}$ and CH $r_{-}$. Form eq. (168), we find that $R$ naively diverges at the extreme rotational KBH when $r_{+}=r_{-}$. It changes the extreme condition to $M^{2}=a^{2}-\alpha$, for this case the Hawking temperature vanishes and the thermodynamics will break down. This result is different from Kerr black hole. The other divergence of scalar curvature for rotational KBH occurs at
\begin{equation}
[\alpha^{2}+(r_{+}+r_{-})^{2}(\alpha+r_{+}r_{-})][(r_{+}+r_{-})^{2}(6r^{2}_{+}+3\alpha+3r_{+}r_{-})+
\alpha(r_{+}+r_{-})r_{+}+\alpha^{2}]-(r^{2}_{+}+\alpha+$$$$
r_{+}r_{-})^{4}-2r^{2}_{-}=0.
\label{168}
\end{equation}
Comparing with eq. (68) and eq. (\ref{142}), We found that they are self-consistent. The heat capacity also becomes singular under fixed angular momentum. Due to the existence of radiation, the phase transition have been changed.

\section{SUMMARY}
\label{sum}
In this work, we have discussed the thermodynamic properties and phase transition for rotational KBH in semi-classical method. Through horizon perturbation, we calculate the thermodynamic quantity and its products. For quintessence matter, radiation and dust case, we had a detailed discussion. We found that these thermodynamic properties depend on the equation of state $\omega$ and strength parameter $\alpha$. We also generalize the Smarr mass formula and Christodoulou-Ruffini mass formula, and find that the first law of thermodynamics is satisfied. Finally we calculate the phase transition of the rotational KBH with radiation case $\omega=1/3$, and find that this phase transition is the second-order phase transition. By the thermodynamic geometry method, we obtain the expression of scalar curvature. For the radiation around Kerr black hole, the phase transition condition and thermodynamic critical phenomena would be changed.

\acknowledgments
We acknowledge the anonymous referee for a constructive comments. We acknowledge the financial support from the National Natural Science Foundation of China 11573060, 11661161010.

\end{document}